\documentclass[aip,reprint,letterpaper,showpacs,amsmath,amssymb, superscriptaddress]{revtex4-2}

\draft % marks overfull lines with a black rule on the right

\usepackage{graphicx}
\usepackage{subfigure}
\usepackage[T1,T2A]{fontenc}
\usepackage[english]{babel}

\usepackage{color}

\newcommand{\delV}[1]{\textcolor{red}{\bf }}

\newcommand{\delMA}[1]{\textcolor{red}{ }}
\usepackage[normalem]{ulem}

\begin{document}

%\preprint{}

\title{Spatiotemporal dynamics of free and bound excitons in CVD-grown MoS$_2$ monolayer} %Title of paper



%\author{}
%\thanks{}
%\affiliation{}


\author{M. A. Akmaev}
%\email[]{akmaevma@lebedev.ru}
\thanks{Authors to whom correspondence should be addressed: akmaevma@lebedev.ru; belykh@lebedev.ru}
\affiliation{P.N. Lebedev Physical Institute of the Russian Academy of Sciences, Moscow, 119991 Russia}

\author{M. M. Glazov}
\affiliation{Ioffe Institute, Russian Academy of Sciences, St. Petersburg, 194021 Russia}

\author{M. V. Kochiev}
\affiliation{P.N. Lebedev Physical Institute of the Russian Academy of Sciences, Moscow, 119991 Russia}

\author{P. V. Vinokurov}
\affiliation{M.K. Ammosov North-Eastern Federal University, Yakutsk, 677000 Russia}

\author{S. A. Smagulova}
\affiliation{M.K. Ammosov North-Eastern Federal University, Yakutsk, 677000 Russia}

\author{V. V. Belykh}
%\email[]{belykh@lebedev.ru}
\thanks{Authors to whom correspondence should be addressed: akmaevma@lebedev.ru; belykh@lebedev.ru}
\affiliation{P.N. Lebedev Physical Institute of the Russian Academy of Sciences, Moscow, 119991 Russia}

\date{\today}

\begin{abstract}
We study photoluminescence (PL) spectra and exciton dynamics of MoS$_2$ monolayer (ML) grown by the chemical vapor deposition technique. In addition to the usual direct A-exciton line we observe a low-energy line of bound excitons dominating the PL spectra at low temperatures. This line shows unusually strong redshift with increase in the temperature and submicrosecond time dynamics suggesting indirect nature of the corresponding transition. By monitoring temporal dynamics of exciton PL distribution in the ML plane we observe diffusive transport of A-excitons and measure the diffusion coefficient up to $40$~cm$^2$/s at elevated excitation powers. The bound exciton spatial distribution spreads over tens of microns in $\sim 1$ $\mu$s. However this spread is subdiffusive, characterized by a significant slowing down with time. The experimental findings are interpreted as a result of the interplay between the diffusion and Auger recombination of excitons.
\end{abstract}


\maketitle %\maketitle must follow title, authors, abstract and \pacs

%\section{Introduction}
Atomically thin transition metal dichalcogenides (TMDC) layers offer a broad tunability of their optical and electronic properties and are considered as a promising replacement for conventional semiconductor structures in electronics and optoelectronics 
\cite{Wang_Electronics_2012, Geim_Van_2013, Liu_Van_2016, Mak_Photonics_2016, Choi_Recent_2017, Manzeli_2D_2017, Durnev_Excitons_2018, Chernozatonskii_Quasitwodimensional_2018, Wang_Colloquium_2018, Glazov_Valley_2021}. As a result of strong electron confinement two-dimensional (2D) TMDC exhibit a number of unique features compared to the traditional semiconductor quantum wells. Among them are the exciton binding energy of the order of several hundreds of meV \cite{Durnev_Excitons_2018, Cheiwchanchamnangij_Quasiparticle_2012, Ramasubramaniam_Large_2012, Chernikov_Exciton_2014, Wang_Colloquium_2018}, strong spin-orbit interaction and spin-valley locking \cite{Zhu_Giant_2011, Glazov_Spin_2015, Wang_Colloquium_2018}, strong dependence of the band structure on the number of layers \cite{Mak_Atomically_2010, Cheiwchanchamnangij_Quasiparticle_2012}.  These and other features make TMDC also promising objects for valley- and spintronics \cite{Mak_Control_2012, Zeng_Valley_2012, Mai_ManyBody_2014, Jariwala_Emerging_2014, Glazov_Spin_2015, Schaibley_Valleytronics_2016, Neumann_Optovalleytronic_2017}.

Photoluminescence (PL) spectra of TMDC monolayers are dominated by exciton lines even at room temperature \cite{Mak_Atomically_2010, Splendiani_Emerging_2010, Wang_Colloquium_2018} which stimulates intense studies of exciton physics in these structures. At low temperatures the inhomogeneous potential profile felt by excitons becomes especially important leading to the intense PL lines of bound excitons. The nature of the bound exciton states is determined by the effect of a substrate, quality of the bulk material and monolayer preparation technique \cite{Amani_Nearunity_2015, Cadiz_Well_2016, Lin_Defect_2016, Hong_Atomic_2017, Wu_Spectroscopic_2017, Hu_Twodimensional_2018, Zhao_Recent_2019, Zhou_How_2021}. Defect signatures in TMDC monolayers have been observed since the very first works \cite{Korn_Lowtemperature_2011, Plechinger_Lowtemperature_2012, Tongay_Defects_2013, Zhou_Intrinsic_2013}, and are being actively studied \cite{Lin_Defect_2016, Hong_Atomic_2017, Wu_Spectroscopic_2017, Hu_Twodimensional_2018, Zhao_Recent_2019, Zhou_How_2021}. Bound exciton states often offer features not inherent to free excitons. One example is the possibility of creating quantum dot-like states %s on defects 
for single photon emission \cite{He_Single_2015, Srivastava_Optically_2015, Kumar_StrainInduced_2015, Tonndorf_Singlephoton_2015, Branny_Discrete_2016}. In addition to unintentional intrinsic defects \cite{Korn_Lowtemperature_2011, Plechinger_Lowtemperature_2012, Saigal_Evidence_2016}, 
defects can be created using a variety of methods \cite{Tongay_Defects_2013, Chow_DefectInduced_2015, Wu_Defect_2017}. 

Despite extensive research activity, the PL dynamics of bound excitons in TMDC monolayers is still poorly understood. Moreover, the spatiotemporal dynamics of bound excitons has not been investigated, unlike the diffusion of free excitons which was studied in monolayers of  
WS$_2$ \cite{Yuan_Exciton_2017, Kulig_Exciton_2018, Goodman_SubstrateDependent_2020, Liu_Direct_2020, Rosati_Straindependent_2021, Fu_Effect_2019, Zipfel_Exciton_2020}, 
WSe$_2$ \cite{Cui_Transient_2014, Mouri_Nonlinear_2014, Cadiz_Exciton_2018, CordovillaLeon_Exciton_2018, CordovillaLeon_Hot_2019}, 
MoSe$_2$ \cite{Kumar_Exciton_2014, Hotta_Exciton_2020, Hao_Controlling_2020} and MoS$_2$ \cite{Goodman_SubstrateDependent_2020, Uddin_Neutral_2020, Yu_Giant_2020}.

In this work, we observe the bound exciton states inherent to chemical vapor deposition (CVD) grown MoS$_2$ monolayer which show a number of interesting features, in particular giant redshift with temperature and extremely long PL decay time. We compare optical properties of bound excitons and free A-excitons for various temperatures and excitation powers and the PL dynamics in time and space. We directly observe the expansion of the spatial distribution of A-excitons at room temperature, which has a diffusive character, and the dynamics of spatial distribution with time of bound excitons at helium temperatures, which is much slower. We discuss the mechanisms driving the spatiotemporal dynamics of bound excitons and highlight the importance of the Auger processes. 

\begin{figure}
\begin{center}
\includegraphics[width=1\columnwidth]{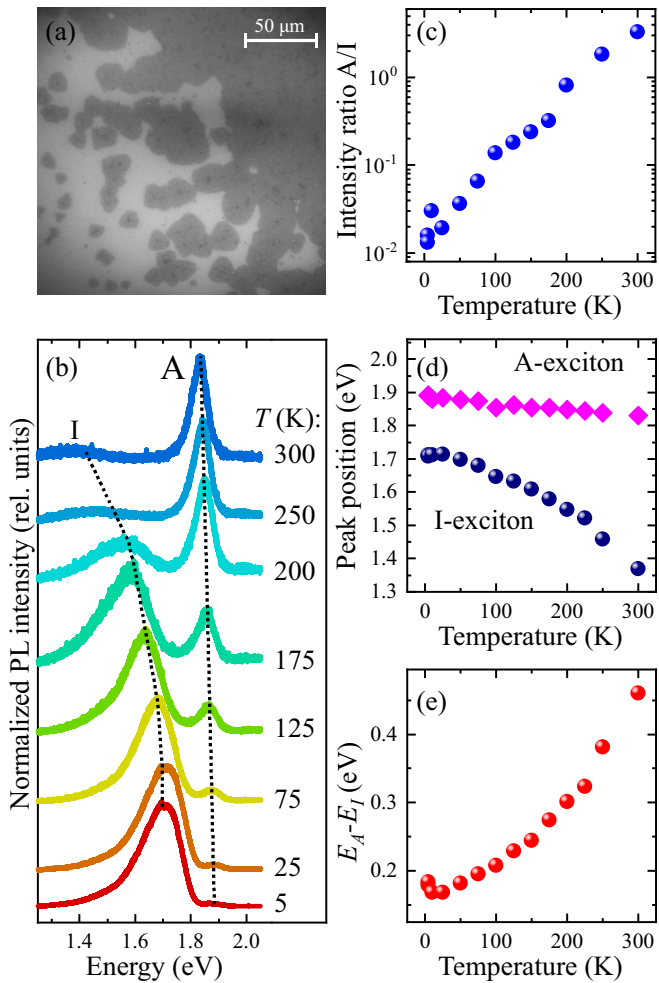}
\caption{(a) Optical image of the CVD-grown MoS$_2$ monolayer. (b) PL spectra of the MoS$_2$ monolayer for different temperatures. ``A'' denotes the direct A-exciton, ``I'' is the bound exciton. The spectra are normalized to their maxima. (c) Temperature dependence of the ratio of intensities of the A- and I-excitons. (d) Temperature dependence of the energy peak position of the A- and  I-excitons. (e) Temperature dependence of the difference between positions of the A- and  I-excitons.}
\label{fig:PL}
\end{center}
\end{figure}

%\section{Experimental details}
The studied structure is a CVD-grown MoS$_2$ monolayer on a Si/SiO$_2$ substrate. Sample growth technique is presented in Ref.~\onlinecite{Smagulova_Study_2020}. Figure~\ref{fig:PL}(a) shows an optical image of the structure. The image was taken at the boundary of a continuous monolayer film. The islands are separate domains of a MoS$_2$ monolayer with a lateral size of several tens of $\mu$m increasing when going to the right where they form a continuous monolayer film of several millimeters in size. Black dots in the center of the islands are the nucleation centers of the multilayer.The presence of monolayer regions was also confirmed by Raman spectroscopy and AFM measurements \cite{Smagulova_Study_2020}.

For PL measurements the sample was placed in a helium-flow cryostat to achieve the temperatures 5$\div$400~K. To focus laser emission on the sample and to collect PL the micro-objective is used. The diameter of the laser spot on the sample is up to 3 $\mu$m. 

The steady-state PL spectra are recorded with a resolution of 0.5 meV using a monochromator coupled to a cooled Si CCD matrix, while the sample was excited using a CW semiconductor laser with a wavelength of 457 nm.

The PL dynamics is measured with a Hamamatsu streak camera coupled to a monochromator. In this case, the sample is excited at the wavelength of 400 nm by the second harmonic of a pulses from Ti:sapphire laser having a pulse duration of 2~ps and pulse repetition rate of 76~MHz. A pulse picker was used to reduce the pulse repetition rate for the dynamics measurements in a wide time range. A temporal resolution up to 3 ps was achieved. 

%\section{Results}

%\subsection{Steady-state PL}

Studies of the PL spectra confirm the presence of monolayer regions. Figure~\ref{fig:PL}(b) shows the PL spectra of MoS$_2$ at different temperatures. At room temperature  the line from the direct A-exciton ($\sim$ 1.830 eV) is observed. The position of the line coincides with the literature data for CVD-grown MoS$_2$ \cite{Chae_Substrateinduced_2017,Plechinger_Direct_2014} and is about 50~meV lower than that for exfoliated monolayers \cite{Mak_Atomically_2010, Korn_Lowtemperature_2011,McCreary2016}. At different points of the sample, the position of the line varies within 10$\div$15 meV. Apparently, this is due to the presence of small multilayer areas and inhomogeneity of the environment (especially substrate). As the temperature is decreased, a lower energy line (I-exciton) appears [Fig.~\ref{fig:PL}(b)] and dominates in the spectrum at low temperatures. We assign this line to bound excitons \cite{Korn_Lowtemperature_2011, Chow_DefectInduced_2015, Tongay_Defects_2013, Saigal_Evidence_2016}. Below we show that bound excitons may be indirect in the $\bf k$-space. At $T = 5$~K, the position of the I-exciton is $1.725 \pm 0.015$ eV. Figure~\ref{fig:PL}(c) shows the temperature dependence of the ratio of the PL intensity of the A-exciton to that of the I-exciton. The ratio increases more than two orders of magnitude as the temperature increases from 5 to 300~K. Surprisingly, this increase is almost exponential. Possibly, this is related to the activation of excitons from bound I-states to the higher energy states and also increased diffusion towards the traps with rising the temperature, resulting in the decrease of the bound exciton lifetime cf. Ref.~\onlinecite{Zipfel_Exciton_2020}. Also, the energy position of the I-exciton shifts much faster than the position of the A-exciton line with temperature [Fig.~\ref{fig:PL}(d)]. 
Hence, the energy difference between A and I excitons increases from 170 to 450~meV as the temperature is increased from 4 to 300~K [Fig.~\ref{fig:PL}(e)]. 
 
\begin{figure}
\begin{center}
\includegraphics[width=1\columnwidth]{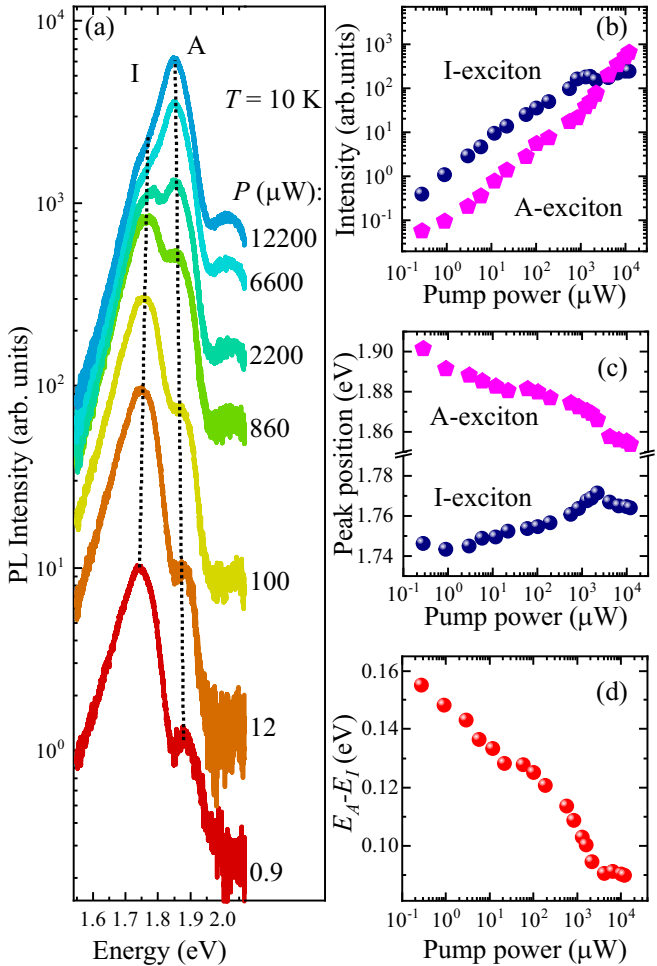}
\label{power}
\caption{(a) PL spectra of MoS$_2$ monolayer for different excitation powers at 10 K. (b) Power dependencies of PL intensities. Sub-linear behavior with saturation for the I-exciton and linear behavior for the A-exciton are observed. (c) Power dependencies of the peak positions. Blueshift for the I-exciton and redshift for the A-exciton are observed. (d) Power dependence of the difference between energies of the A-exciton and the I-exciton. }
\label{fig:power}
\end{center}
\end{figure}

Figure~\ref{fig:power}(a) shows the PL spectra for different pump powers at $T=10$~K. With increasing the pump power [Fig.~\ref{fig:power}(c)], a red shift of the A-exciton line is observed, which can be related to exciton interaction effects and to local overheating. On the other hand, the I-exciton line shows a blue shift. This may be related to the fact that I-excitons are localized in traps which are characterized by several energy levels. As the pump power is increased, the higher energy levels of the traps are filled and, as a consequence, the emission line shifts to higher energies. One can see from Fig.~\ref{fig:power}(b), that the I-exciton line intensity has a sub-linear dependence on the pump power and saturates at high powers. At the same time, the A-exciton line intensity has a linear power dependence and dominates at high powers. These observations also indicate that the I-exciton line corresponds to excitons bound on defects or impurities. Note, that at the excitation powers where the I-exciton intensity saturates, its energy position changes the trend from increase with the increasing the power to decrease. This indicates, that for pump powers corresponding to the saturation, energy shifts of the A-exciton and the I-exciton lines have the same nature. Note that the sample does not degrade as a result of high power excitation, which is confirmed by reproducibility of the measurements carried out before and after the high power excitation. Additional confirmation of the bound nature of I-excitons comes from significant temperature-induced redshift of I-line, Figs.~\ref{fig:PL}(b,d), which is most probably related to the thermal activation of localized excitons and its subsequent trapping to the lower energy states.\cite{tar01}

%\subsection{Dynamics of excitons}

\begin{figure}
\begin{center}
\includegraphics[width=0.95\columnwidth]{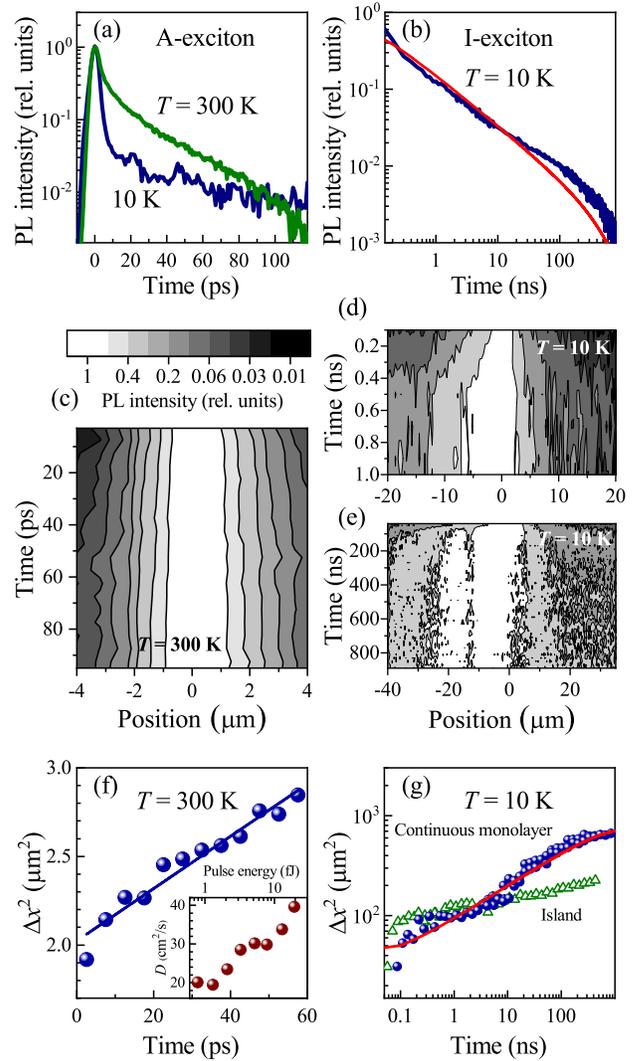}
\caption{(a) PL dynamics of MoS$_2$ monolayer in a short time range corresponding to the A-exciton for low and high temperatures. (b) PL dynamics of MoS$_2$ monolayer in a broad time range corresponding to the I-exciton. Red line shows the result of theoretical calculation involving Auger processes. (c --- e) Dynamics of PL spatial profiles in different time ranges. The spatial distribution at each time moment is normalized to maximum. The shortest time range at $T = 300$~K corresponds to the diffusion of A-excitons, while broadening of the PL distribution at longer times at $T=10$~K corresponds to the transport of I-excitons. (f) Squared width of the PL spatial distribution $\Delta x^2$ as a function of time for A-excitons at $T=300$~K and excitation pulse energy 13~fJ. The dynamics shows diffusive character $\Delta x^2(t) = \Delta x^2(0) + 4Dt$. The inset shows the power dependence of the diffusion coefficient $D$. (g)~Squared width of PL spatial distribution $\Delta x^2$ as a function of time for I-excitons at $T=10$~K. The dynamics is subdiffusive. Open triangles show the corresponding dynamics for excitons confined within monolayer island. Red line shows the result of theoretical calculation involving Auger processes.}
\label{fig:diffusion}
\end{center}
\end{figure}

Next we study temporal dynamics of the PL from the A and I exciton lines. The dynamics on a short time scale of about 100~ps [Fig.~\ref{fig:diffusion}(a)] is dominated by A-exciton (see the Supplemental material). We observe two components of the decay: fast and slow. The fast component has the decay time within our temporal resolution of 3~ps, while the characteristic decay of the slow component is several tens of picoseconds. As the temperature is decreased, the amplitude of the slow component decreases with respect to that of the fast one, while its decay time increases, Fig.~\ref{fig:diffusion}(a) and Supplemental material. A similar decrease of the slow component amplitude was observed in Ref.~\onlinecite{Akmaev_Nonexponential_2020} on a considerably longer timescale and was attributed to the inhomogeneous energy dependent distribution of exciton recombination times. While PL from A-excitons decays relatively fast, emission from I-excitons at low temperature is visible on the much longer timescale up to 1 $\mu$s [Fig.~\ref{fig:diffusion}(b)]. Its decay is slow and nonexponential. From 1 to 100~ns the decay is close to the power-law dependence, while at longer times it accelerates. Thus, trapping of excitons suppresses their radiative recombination. This is in agreement with previous reports on long bound exciton lifetimes \cite{Cadiz_Well_2016,Goodman2017}. 

The long dynamics suggests an indirect transition nature of this I-exciton line. The CVD growth technique results in a large number of inhomogeneities and defects. They cause strain, which leads to a direct-indirect band gap transition \cite{Ma_Direct_2020, Blundo_Evidence_2020, Peto_Moderate_2019, Chaste_Intrinsic_2018, Fu_KL_2017, Conley_Bandgap_2013, Yue_Mechanical_2012, Eliseyev_Photoluminescence_2021}. Excitons bound to these defects may also have an indirect nature.

%\subsection{Diffusion}

Finally, we study the transport of the A-excitons and I-excitons by monitoring the dynamics of the PL spatial distribution. This distribution normalized to the maximum intensity at each time moment is shown in Fig.~\ref{fig:diffusion}(c) for A-exciton at $T = 300$~K. The corresponding squared width $\Delta x$ of this distribution as a function of time is shown in Fig.~\ref{fig:diffusion}(f). The width $\Delta x$ is obtained from fitting the spatial PL distribution with Gaussian function $\exp{[-(x-x_0)^2/\Delta x^2]}$. It shows a linear increase with time $\Delta x^2(t) = \Delta x^2(0) + 4Dt$ indicating diffusive transport of A excitons at room temperature with the diffusion coefficient $D$ increasing with excitation power from 20 to 40 cm$^2$/s [inset in Fig.~\ref{fig:diffusion}(f)]. The increase of the diffusion coefficient with excitation density was also reported for WS$_2$ monolayers \cite{Kulig_Exciton_2018}, for WSe$_2$ monolayers~\cite{CordovillaLeon_Hot_2019}, and  for MoS$_2$ monolayers \cite{Uddin_Neutral_2020} and is described as a result of the Auger recombination which flattens the exciton distribution and makes their observed diffusion coefficient larger\cite{Kulig_Exciton_2018}. At low temperatures the diffusion of A-exciton is suppressed and hardly can be detected within their lifetime. On the contrary, diffusion of I-excitons can be studied at low temperatures where their PL is considerable. The slowness of the diffusion here is compensated by the slow decay of the PL. Figures~\ref{fig:diffusion}(d),(e) show the spatiotemporal PL dynamics for the I-exciton in different time ranges at $T = 10$~K. The PL distribution apparently broadens with time, while its center shifts. The last effect may be realted to the inhomogeneous potential profile or inhomogeneous defect distribution. Figure~\ref{fig:diffusion}(g) shows time dynamics of $\Delta x^2$ for I-excitons over 1~$\mu$s time range for excitation of the continuous monolayer region (full balls) and island of 10-$\mu$m-size (open triangles). The exciton transport within the island is limited by its size. On the other hand, I-excitons created in continuous monolayer expand their distribution over several tens of microns. Nevertheless, time dependence of $\Delta x^2$ is subdiffusive. We have analyzed possible origins of the effect, see the Supplementary material for details, including anomalous diffusion with large spread of hopping times, interplay of diffusion and energy relaxation and the Auger effect. It turns out that accounting for the exciton-exciton non-radiative recombination allows us to reproduce the key features of the PL dynamics, Fig.~\ref{fig:diffusion}(b) and $\Delta x^2(t)$ dependence, Fig.~\ref{fig:diffusion}(g), with reasonable set of parameters of the linear diffusion coefficient $D_0 = 0.003$~cm$^2$/s and the Auger rate $0.045$~$\mu$m$^2$/ns, see Supplementary material for details.

See supplementary materials for details of PL dynamics and of model of subdiffusive dynamics of I-excitons.

\begin{acknowledgements}
We are grateful to M.L. Skorikov for fruitful discussions and valuable advice. 

The work was supported by the Government of the Russian Federation (Contract No. 075-15-2021-598 at the P.N. Lebedev Physical Institute). M.M. Glazov was supported by the RFBR grant No. 19-02-00095. The work on the sample growth was supported by the Project FSRG-2020-0017 of the state assignment of the Ministry of Education and Science of Russia for 2020-2022
\end{acknowledgements}

\section*{Data Availability Statement}
The data that support the findings of this study are available from the corresponding author upon reasonable request.

%\section{}
%\label{}
%\subsection{}
%\subsubsection{}

% If in two-column mode, this environment will change to single-column format so that long equations can be displayed. 
%\begin{widetext}
%$$\mbox{put long equation here}$$
%\end{widetext}

% Figures should be put into the text as floats. 
% Use the graphics or graphicx packages (distributed with LaTeX2e).
% Put the label that you will use with \ref{} command in the braces of the \label{} command.
%
% \begin{figure}
% \end{figure}

% \begin{table}
% \end{tabular}
% \end{table}

%\begin{acknowledgments}
% Put your acknowledgments here.
%\end{acknowledgments}

%\bibliography{CVDMoS2}

%\bibliography{AkmaevCVDMoS2-2021}

\end{document}

\title{Supplementary material: Spatiotemporal dynamics of free and bound excitons in CVD-grown MoS$_2$ monolayer}

\author{M. A. Akmaev}
\affiliation{P.N. Lebedev Physical Institute of the Russian Academy of Sciences, Moscow, 119991 Russia}

\author{M. M. Glazov}
\affiliation{Ioffe Institute, Russian Academy of Sciences, St. Petersburg, 194021 Russia}

\author{M. V. Kochiev}
\affiliation{P.N. Lebedev Physical Institute of the Russian Academy of Sciences, Moscow, 119991 Russia}

\author{P. V. Vinokurov}
\affiliation{M.K. Ammosov North-Eastern Federal University, Yakutsk, 677000 Russia}

\author{S. A. Smagulova}
\affiliation{M.K. Ammosov North-Eastern Federal University, Yakutsk, 677000 Russia}

\author{V. V. Belykh}
\affiliation{P.N. Lebedev Physical Institute of the Russian Academy of Sciences, Moscow, 119991 Russia}

\maketitle
\setcounter{equation}{0}
\setcounter{figure}{0}
\setcounter{table}{0}
\setcounter{section}{0}
\setcounter{page}{1}
    %\makeatletter
\renewcommand{\theequation}{S\arabic{equation}}
\renewcommand{\thefigure}{S\arabic{figure}}
\renewcommand{\thetable}{S\arabic{table}}
\renewcommand{\thesection}{S\arabic{section}}

\section{Photoluminescence dynamics of free and bound excitons}
\begin{figure}
\begin{center}
\includegraphics[width=0.8\columnwidth]{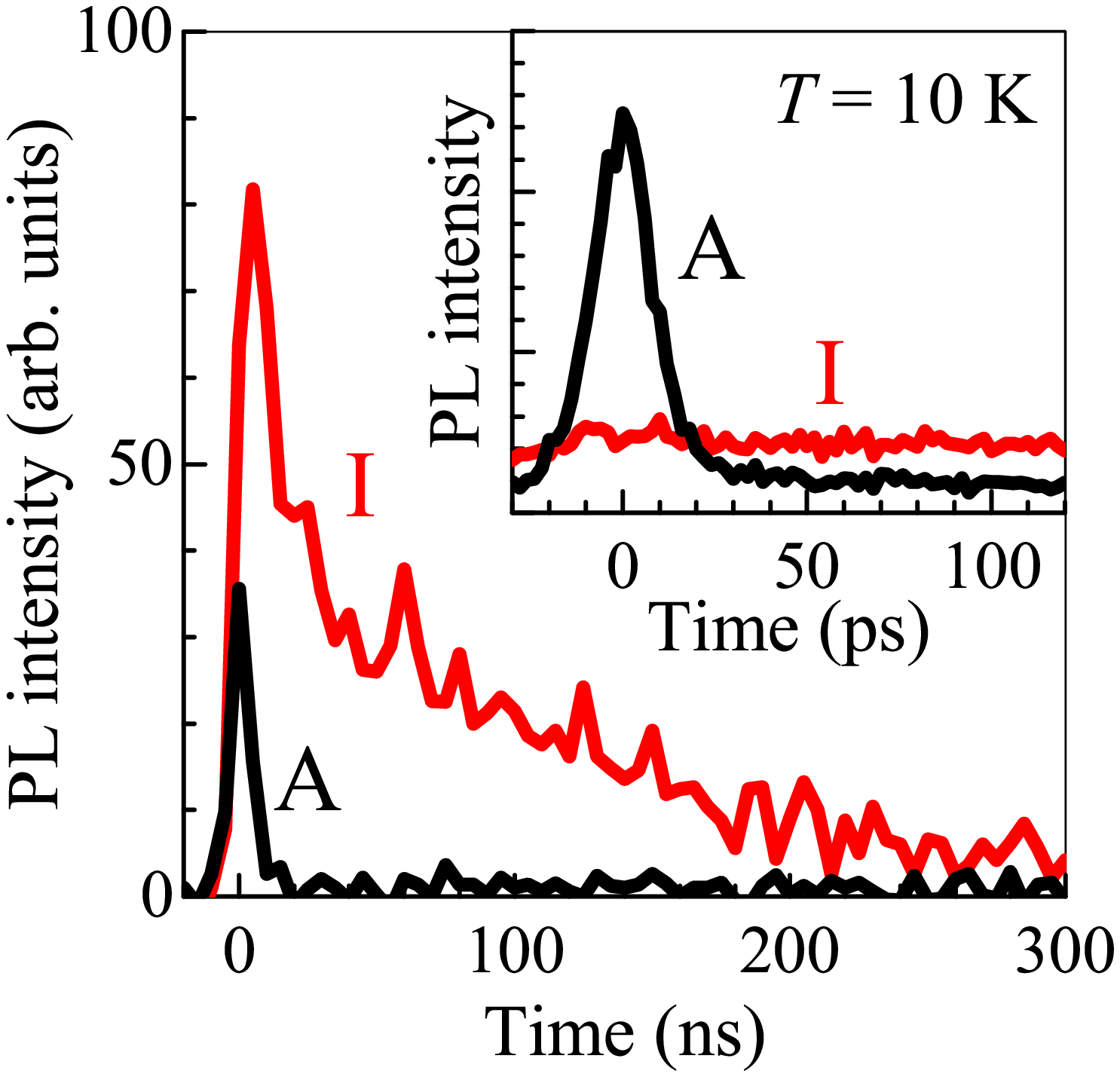}
\caption{PL dynamics registered at the energies of 1.88~eV (black curve) and 1.73~eV (red curve), corresponding to the spectral positions of free A- and bound I-excitons, respectively. The inset shows the corresponding dynamics in the short time range. The temperature $T = 10$~K.}
\label{fig:WLDep}
\end{center}
\end{figure}
To study the dynamics of photoluminescence (PL) spatial distribution which is presented in the main article, the PL spot was imaged directly to the streak camera slit without spectral separation. Nevertheless we attribute the PL within a 100-ps time range to the free A-exciton, while at longer times the PL mainly corresponds to the bound I-exciton. We show this by performing PL dynamics measurements with spectral resolution (Fig.~\ref{fig:WLDep}). Indeed, the PL dynamics at the energy of 1.88~eV, corresponding to the spectral position of the A-exciton, is within the temporal resolution, while the PL dynamics at 1.73~eV, corresponding to the spectral position of the I-exciton, is 100-ns-slow. This conclusion is confirmed by measurements with increased temporal resolution  (inset in Fig.~\ref{fig:WLDep}). Note, that peak intensity of the A-exciton at very short time is higher than that of the  I-exciton, while the time integrated intensity of the A-exciton is much lower than that for the I-exciton at low temperatures.  

\section{PL dynamics at different temperatures}
\begin{figure}
\begin{center}
\includegraphics[width=1\columnwidth]{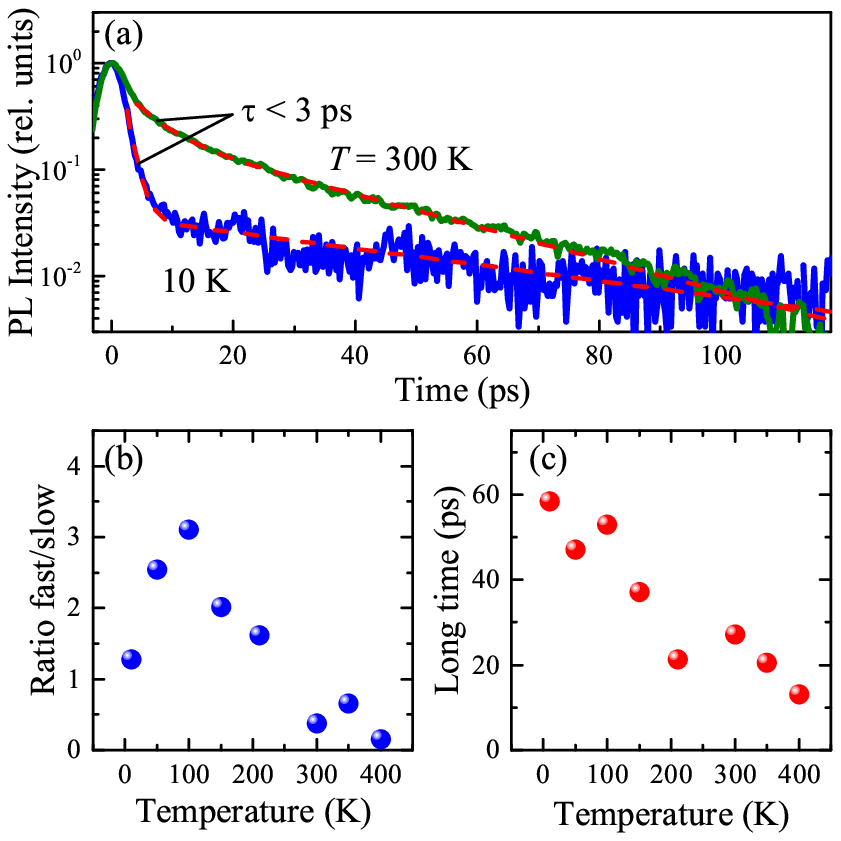}
\caption{(a) PL dynamics at low and high temperatures. Red dashed lines show the double exponential fits. (b) Temperature dependence of the intensity ratio of the fast and slow components in the PL dynamics. (c) Temperature dependence of the slow PL component decay time. The laser pulse energy $W = 20$~fJ.}
\label{fig:TDep}
\end{center}
\end{figure}
The inset in Fig.~\ref{fig:WLDep} shows that at $T = 10$~K the A-exciton PL decay time is rather short, and already at $t > 20$~ps the I-exciton emission dominates. Measurements without spectral resolution but with high temporal resolution reveal the A-exciton decay within 3~ps at $T=10$~K [Fig.~\ref{fig:TDep}(a)]. With increasing temperature in the PL kinetics a slower component appears with characteristic time of several tens of ps. Note that I-exciton emission decreases with increasing temperature (Fig.~1 in the main article). Thus the slower component at higher temperatures (and especially at room temperature) corresponds to A-excitons having wave vectors beyond the light cone. The relative contribution of slower component increases with temperature, while its decay time decreases [Figs.~\ref{fig:TDep}(b) and (c)].     

\section{PL dynamics at different excitation powers}
\begin{figure}
\begin{center}
\includegraphics[width=1\columnwidth]{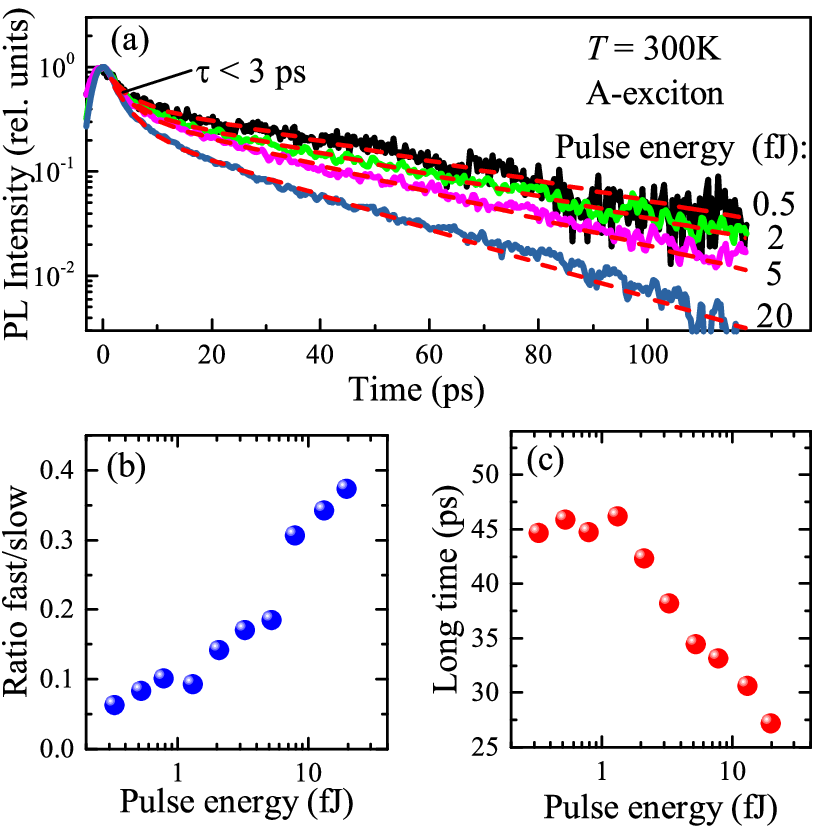}
\caption{(a) PL dynamics of the A-exciton, corresponding to different laser pulse energies. Red dashed lines show the double exponential fits. (b) Pulse energy dependence of the intensity ratio of the fast and slow components in the PL dynamics. (c) Pulse energy dependence of the slow PL component decay time. $T = 300$~K.}
\label{fig:PDepA}
\end{center}
\end{figure}

\begin{figure}
\begin{center}
\includegraphics[width=0.8\columnwidth]{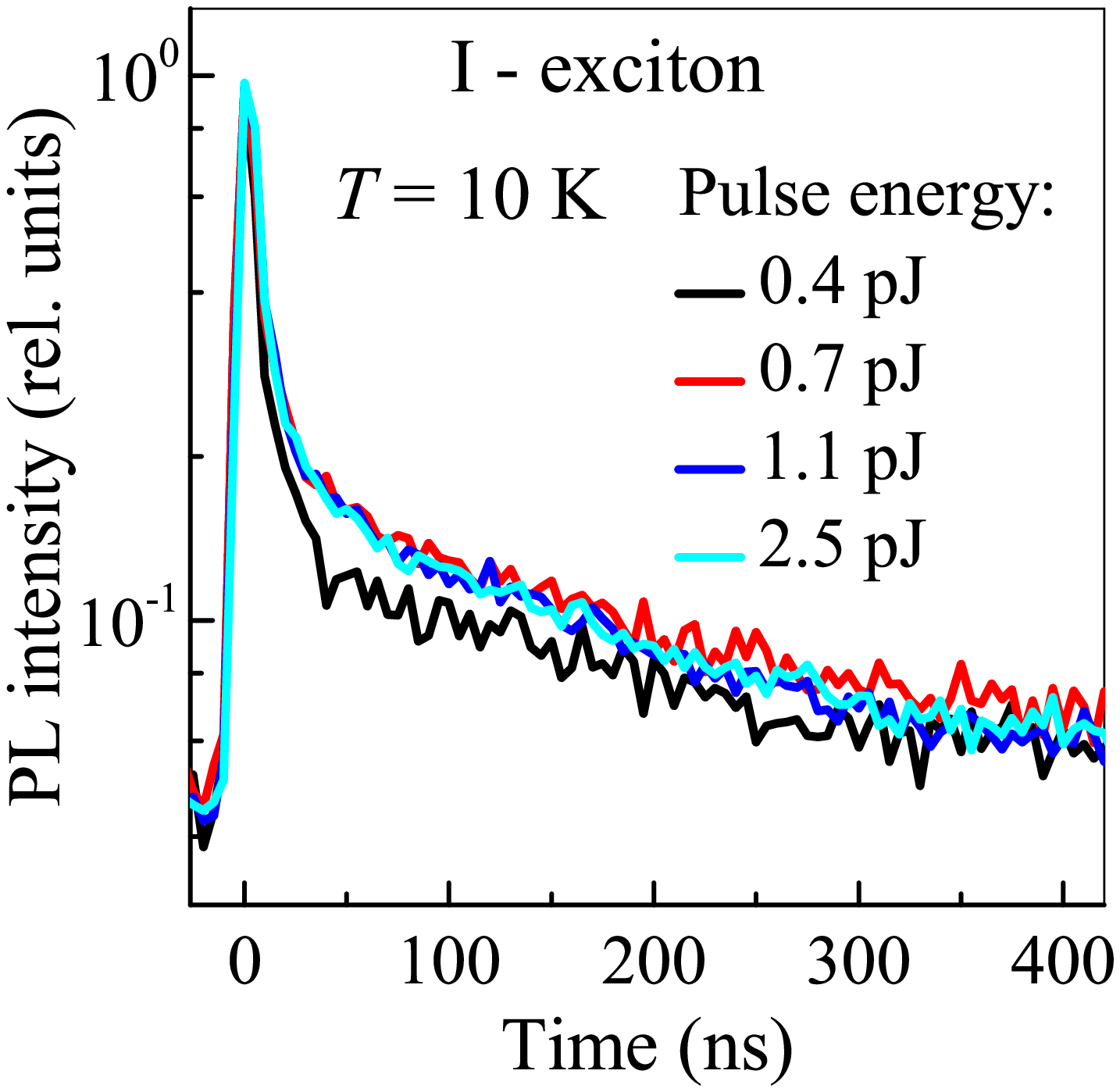}
\caption{(a) PL dynamics of I-exciton, corresponding to different laser pulse energies. $T = 10$~K.}
\label{fig:PDepI}
\end{center}
\end{figure}

The ratio of amplitudes of the short and long components in the PL dynamics of A-exciton increases with increase of the excitation power, characterized by the laser pulse energy $W$ [Fig.~\ref{fig:PDepA}(a),(b)], while the decay time
of the long component decreases [Fig.~\ref{fig:PDepA}(c)]. The overall dynamics shortens with excitation power indicating presence of nonlinear processes. 
The shape of PL dynamics of I-exciton measured at a long timescale is less sensitive to the excitation power (Fig.~\ref{fig:PDepI}).

\section{Model of subdiffusive dynamics of I-excitons}

The transport of I-excitons can be considered as their hopping between the localization sites. There are two main parameters which control the transport in the linear, i.e., low excitation density regime: the jump waiting time $\tau$ and  jump length $l$. For randomly distributed localization sites both $\tau$ and $l$ can have exponentially large spread\cite{efrosshklovskii}. The diffusion coefficient can be estimated as 
\begin{equation}
\label{d:est}
D \sim \left\langle\frac{l^2}{\tau} \right\rangle,
\end{equation}
where the angular brackets denote an appropriate average (strictly speaking, one has to average the logarithm of the diffusion coefficient).

Experiments presented in the main text indicate subdiffusive expansion of the I-excitons. Below we discuss possible mechanisms of the effect.

\subsection{Random walk with divergent waiting time}

 Subdiffusive expansion can occur provided that the characteristic waiting time $\langle \tau\rangle$ diverges. For instance, for the power-law distribution of the waiting times 
\begin{equation}
\label{distrib:tau}
p(\tau) \sim \frac{\tau_0^\alpha}{\tau^{\alpha+1}},
\end{equation}
with $\tau_0$ being the time-scale of the hopping process and $0< \alpha< 1$ results in the fractional diffusion equation\cite{metzler}
\begin{equation}
\label{frac:diff}
\frac{\partial n(\bm \rho,t)}{\partial t} = D_\alpha \Delta \tilde n(\bm \rho,t),
\end{equation}
where $n(\bm \rho,t)$ is the exciton density as a function of the in-plane position $\bm \rho$ and time $t$, $ D_\alpha \sim \langle l^2\rangle/\tau_0^\alpha$ is the generalized diffusion coefficient, $\Delta$ is the two-dimensional Laplace operator, and $\tilde n( \bm\rho,t)$ is defined by
\begin{equation}
\label{tilde:n}
\tilde n(\bm \rho,t) = \frac{1}{\Gamma(\alpha)} \frac{\partial }{\partial t}\int_0^t \frac{n(\bm \rho,t') dt'}{(t-t')^{1-\alpha}},
\end{equation}
with ${\Gamma(\alpha)}$ being the Gamma-function. It follows from Eqs.~\eqref{frac:diff} and \eqref{tilde:n} that the squared displacement 
\begin{equation}
\label{dx:alpha}
\Delta x^2(t) \propto t^\alpha,
\end{equation}
and the exciton expansion is subdiffusive for $\alpha<1$ and diffusive for $\alpha=1$. While this scenario cannot be excluded, the fit of the experimental data gives unrealistically large values of $D_\alpha$: The expansion happens \emph{significantly faster} than for the A-exciton, which contradicts to the assumption of strong localization of I-excitons.

\subsection{Interplay of diffusion and thermalization}

If the I-exciton diffusion takes place via the variable-range hopping the diffusion coefficient is a function of the exciton temperature $T$ (which typically exceeds the lattice temperature) as~\cite{efrosshklovskii}
\begin{equation}
\label{DT}
D= D_0 \exp{\left[-\left(\frac{T_0}{T}\right)^{1/3}\right]}.
\end{equation}
Here we took into account that the exciton motion is possible in the two-dimensional plane only, $T_0$ and $D_0$ are the parameters related to the details of the disorder.

In the course of time the energy of the non-equilibrium exciton ensemble relaxes and the temperature $T$ becomes a function of time, as a result $D\equiv D(t)$. In the simplest possible model we assume that the temperature of excitons decays exponentially as
\begin{equation}
\label{Tt}
T(t) = T(0) \exp{(-t/\tau_\epsilon)},
\end{equation}
where $\tau_\epsilon$ is the energy relaxation time (note that the energy relaxation of the localized excitons is quite complicated and cannot be generally described by a single time $\tau_\epsilon$,~\cite{Golub:1996aa,Tarasenko:2001aa,Glazov:2018tf} we disregard these effects here). We assumed also that the lattice temperature $T_l$ is much lower than $T_0$ making it possible to neglect $T_l$ in Eq.~\eqref{Tt}.

It follows than that the squared width of the exciton distribution $\Delta x^2(t)$ obeys the equation\cite{PhysRevLett.120.207401}
\begin{multline}
\label{w2}
\Delta x^2(t)=\Delta x^2(0) + 4\int_0^t D(t') dt'\\
 = w_0^2 + 4D_0\int_0^t \exp{\left[-\left(\frac{T_0}{T(0)}\right)^{1/3}\exp{\left(\frac{t'}{3\tau_\epsilon}\right)}\right]}dt'.
\end{multline}
Note that this integral can be evaluated through the exponential integral function $\Ei(z) = \int_{-\infty}^x (e^x/x) dx$ as
\begin{multline}
\label{w2:1}
w^2(t)=w_0^2 + 12 D_0 \tau_\epsilon \\
\times \left[\Ei\left(-\sqrt[3]{e^{t/\tau_\epsilon} \frac{T_0}{T(0)}}\right) - \Ei\left(-\sqrt[3]{\frac{T_0}{T(0)}}\right)\right].
\end{multline}
The fit of the experimental data with Eq.~\eqref{w2:1} is possible, but with the initial value of the diffusion coefficient $D_0 \approx 30$~cm$^2$/s, which exceeds the values for the A-exciton at low intensity in contradiction to the assumptions of the strong exciton localization. 

\subsection{Auger-diffusion model}

An interplay of the bimolecular recombination and diffusion results in the increase of the observed diffusion coefficient at short time scales and in the subdiffusive expansion of excitons at the longer times.\cite{PhysRevLett.120.207401} Here we adopt the semi-analytical model developed in Refs.~\onlinecite{PhysRevLett.120.207401,dequilettes2021accurate} where the diffusion equation with account for the Auger effect
\begin{equation}
\label{diff:auger}
\frac{\partial n(\bm \rho,t)}{\partial t} = D_0 \Delta n(\bm \rho,t) - \frac{n(\bm \rho,t)}{\tau} - R_A n^2(\rho,t),
\end{equation}
with monomolecular recombination time $\tau$ and bimolecular recombination rate (Auger rate) $R_A$ is reduced, using the diffusion ansatz 
\begin{multline}
\label{diff:ans}
n(\bm \rho,t)  \\
= \frac{N(t)}{\pi [w_0^2 + 4D_{\rm eff}(t)t]}\exp{\left(-\frac{\rho^2}{w_0^2+4D_{\rm eff}(t)t} \right)},
\end{multline}
to a set of coupled ordinary differential equations for the exciton number $N(t) = \int d\bm \rho n(\bm \rho,t)$ and the effective diffusion coefficient $D_{\rm eff}(t)$ in the form
\begin{subequations}
\label{diff:auger:ans}
\begin{align}
&\frac{dN}{dt} =-\frac{N}{\tau} - \frac{R_A N^2}{2\pi[w_0^2+4D_{\rm eff}(t)t]},\\
&\frac{d(tD_{\rm eff})}{dt} = D_0 + \frac{R_A N}{16\pi}. 
\end{align}
\end{subequations}
Here $D_0$ is the linear diffusion coefficient. We stress that Eq.~\eqref{diff:ans} is a convenient approximation allowing us to obtain the semi-analytical results. The justification and discussion of the validity range of Eq.~\eqref{diff:ans} is presented in Refs.~\cite{PhysRevLett.120.207401,dequilettes2021accurate}.

The set~\eqref{diff:auger:ans} allows to describe simultaneously the emission intensity dynamics and the I-exciton expansion, Fig. 3(b) and (g) of the main text with reasonable set of parameters: $D_0 = 0.003$~cm$^2$/s, $R_A=0.045$~$\mu$m$^2$/ns, $w_0 = 3.5$~$\mu$m, $N(0)\approx 7\times 10^4$. Thus we conclude that the subdiffusive expansion of the I-exciton cloud is mainly related to the interplay between the Auger effect and diffusion.